\documentclass[preprint]{aastex}
\usepackage{natbib}

\begin{document}

\title{Magnetic complexity as an explanation for bimodal rotation populations among young stars}

\author{ Cecilia Garraffo\altaffilmark{1}, Jeremy
  J. Drake\altaffilmark{1},  Ofer Cohen\altaffilmark{1}}

\altaffiltext{1}{Harvard-Smithsonian Center for Astrophysics, 60 Garden St. Cambridge, MA 02138}


\begin{abstract}

Observations of young open clusters have revealed a bimodal
distribution of fast and slower rotation rates that has proven
difficult to explain with predictive models of spin down that depend
on rotation rates alone.  The Metastable Dynamo Model proposed
recently by Brown, employing a stochastic transition probability from
slow to more rapid spin down regimes, appears to be more successful
but lacks a physical basis for such duality.  Using detailed 3D MHD
wind models computed for idealized multipole magnetic fields, we show
that surface magnetic field complexity can provide this basis.  Both
mass and angular momentum losses decline sharply with increasing field
complexity.  Combined with observation evidence for complex field morphologies in magnetically active stars, our results support a picture in which young, rapid rotators lose angular
momentum in an inefficient way because of field complexity.   During this slow spin-down phase, magnetic complexity is eroded, precipitating a rapid transition from weak to strong wind coupling.  
\end{abstract}


\section{INTRODUCTION}
\label{sec:Intro}

Stellar rotation catalyzes magnetic dynamo activity in the interiors of late-type stars that is manifest at the surface in the form of magnetic fields, energetic photon and particle radiation, supersonic winds and coronal mass ejections.  
The magnetized winds carry away angular momentum, a process commonly referred to as ``magnetic braking''.
As stars age, their rotation rates, $\Omega$, eventually converge to the empirical \cite{Skumanich72} spin down law $ \Omega
\sim t^{-1/2}$. 
\citet{Weber67} and \citet{Mestel68} derived the first analytical expression for stellar 
 angular momentum loss, obtaining $\dot{J}=\frac{2}{3} \Omega \dot{M}R_A^2$,
where $\dot{J}$ is the angular momentum loss rate, 
$\dot{M}$ the mass loss rate, $R_A$ is the radial distance at which the wind speed exceeds the local Alfv\'en speed (the ``Alfv\'en radius''), and where a constant radial field was assumed at the surface of the star.
This approach was later generalized
to more realistic scenarios in a range of different studies
\citep[e.g.][]{Mestel87, Kawaler88, Taam89,Chaboyer95a}, which have generally proven successful in explaining spin down on the Main Sequence, including the empirical Skumanich law.

The early phase of rotation evolution has proven more difficult to understand.
Observations of young open clusters by \cite[e.g.][see \citealt{Meibom11} for a recent compilation]{Stauffer87, Soderblom93, Queloz98, Terndrup00} found a large spread in rotation rates at ages up to a few hundred Myrs with a bimodal aspect comprising two branches corresponding to
fast and slow rotation and implying an extremely fast transition between the
two.  
The currently favored explanation for this phenomenon is 
a core-envelope decoupling near the zero age main-sequence \citep[e.g.][]{Stauffer84,Soderblom93,Barnes03}, after which the outer convection zone with lower moment of inertia is rapidly spun down, leaving a more rapidly-rotating core.  
More recently, 
\cite{Brown14} has proposed a ``Metastable Dynamo Model'' (MDM) in which coupling between the magnetic field and wind is initially weak.  Spontaneous strong coupling
of the star to the wind then happens at a certain early
age, initiating the rapid spin down.   For {\it ad hoc} coupling constant changes by factors of 100 or more, the model is successful in reproducing the observed rotation distributions of young clusters.  However, the mechanism behind such a change in coupling has not been identified.   Except for a small
handful of indirect detections \citep[e.g.][]{Wood04, Wood14}, observational progress is stymied by the winds themselves being generally weak, while surface magnetic fields can only be inferred indirectly and with very limited spatial resolution \citep{Donati09}.

One key ingredient in stellar rotation evolution models that has received scant attention is the morphology of the magnetic field.  While the early treatments of \citet{Mestel84} and \citet{Kawaler88} that provided much of the basis for subsequent rotation evolution models considered the multipole order of the magnetic field, this was limited to the effect of the radial dependence of the field strength.  Moreover, spin down models have generally assumed dipolar fields.  Instead, 
a growing database of Zeeman-Doppler imaging observations indicates that surface magnetic fields of young, active stars mainly consist of high-order multipole components, rather than a simple dipole such as characterizes the large-scale solar magnetic field \citep[e.g.][]{Donati03,
Donati09, Marsden11a, Waite11, Waite15}.    \cite{Linsky.Wood:14} have also recently inferred mass loss rates for the active stars $\xi$~Boo~A and $\pi^1$~UMa that are two orders of magnitude lower than expected based on extrapolation from lower activity stars, suggesting that magnetic topology could have a more profound effect on angular momentum loss than simply through the radial field strength dependence.

Here, we investigate the role of magnetic field complexity on stellar
angular momentum loss using a detailed, self-consistent
three-dimensional magnetohydrodynamic (MHD) wind model 
that has proven successful in matching observations of the solar wind \citep{Oran13}. We explore a range of simple magnetic configurations with different multipolar complexity.   
The numerical methods are described in Section~\ref{sec:Methods}, the results of model calculations in 
Section~\ref{sec:Results}, and we discuss our main findings
and their implications in
Section~\ref{sec:Discussion}.  We conclude in Section~\ref{sec:Conclusions} that magnetic complexity can provide the strong coupling switch sought in the MDM by \citet{Brown14}.


\section{NUMERICAL SIMULATION}
\label{sec:Methods}

\subsection{MHD model} 
\label{sec:Model}

In order to obtain solutions for the artificial stellar corona and wind cases considered here, 
we use the generic {\it BATS-R-US} code \citep{Powell99,Toth12} that solves the set of MHD equations for the conservation of mass, momentum, magnetic induction, and energy.  A spherical grid, logarithmic in the $\hat{r}$ coordinate is used with adaptive mesh refinement to resolve current sheets and regions where the magnetic field changes sign. 

The {\it BATS-R-US} module for the solar (or stellar) corona is
driven by synoptic maps of the radial stellar magnetic field, which
are used to specify the field boundary conditions. The initial condition for the three-dimensional field is
obtained by calculating the potential field of these boundary
conditions, assuming that the field is purely radial at a distance of
$r=4.5R_\star$ (the ``source surface") \citep{Altschuler67}. Typically, in the case of the Sun,
the location of the source surface is closer in;  we use
this larger value to prevent the choice of this parameter from having an 
impact on the final solution.  In the case of strong
stellar magnetic fields, choosing a small value for the source surface
can artificially truncate magnetic loops, resulting in an overpowering of the
stellar wind to unrealistic values. This effect is eliminated as the
source surface is moved outwards to larger distances, where it does not impact the MHD steady-state solution. 

Once the initial
condition for the magnetic field is specified, the model provides a
self-consistent acceleration of the wind and heating of the
corona via the Alfv\'en wave turbulence mechanism, including 
thermodynamic processes such as radiative cooling and electron heat
conduction \citep[see][for full
details]{Oran13,Sokolov13,Vanderholst14}.  
Unlike models that were used to study stellar winds with
imposed, fully developed, spherically symmetric thermal (``Parker")
winds \citep[e.g.,][]{Matt12,Vidotto14b}, here the wind and magnetic field solutions evolve together, allowing the magnetic field topology to influence the wind appropriately, as observed in the case of the solar wind in the heliosphere
\citep{Phillips95,McComas07}.   

\subsection{Simulations}

\begin{figure}
\center
\includegraphics[trim = 0.3in 0.3in
  0.5in 3.in,clip, width = 2.12in]{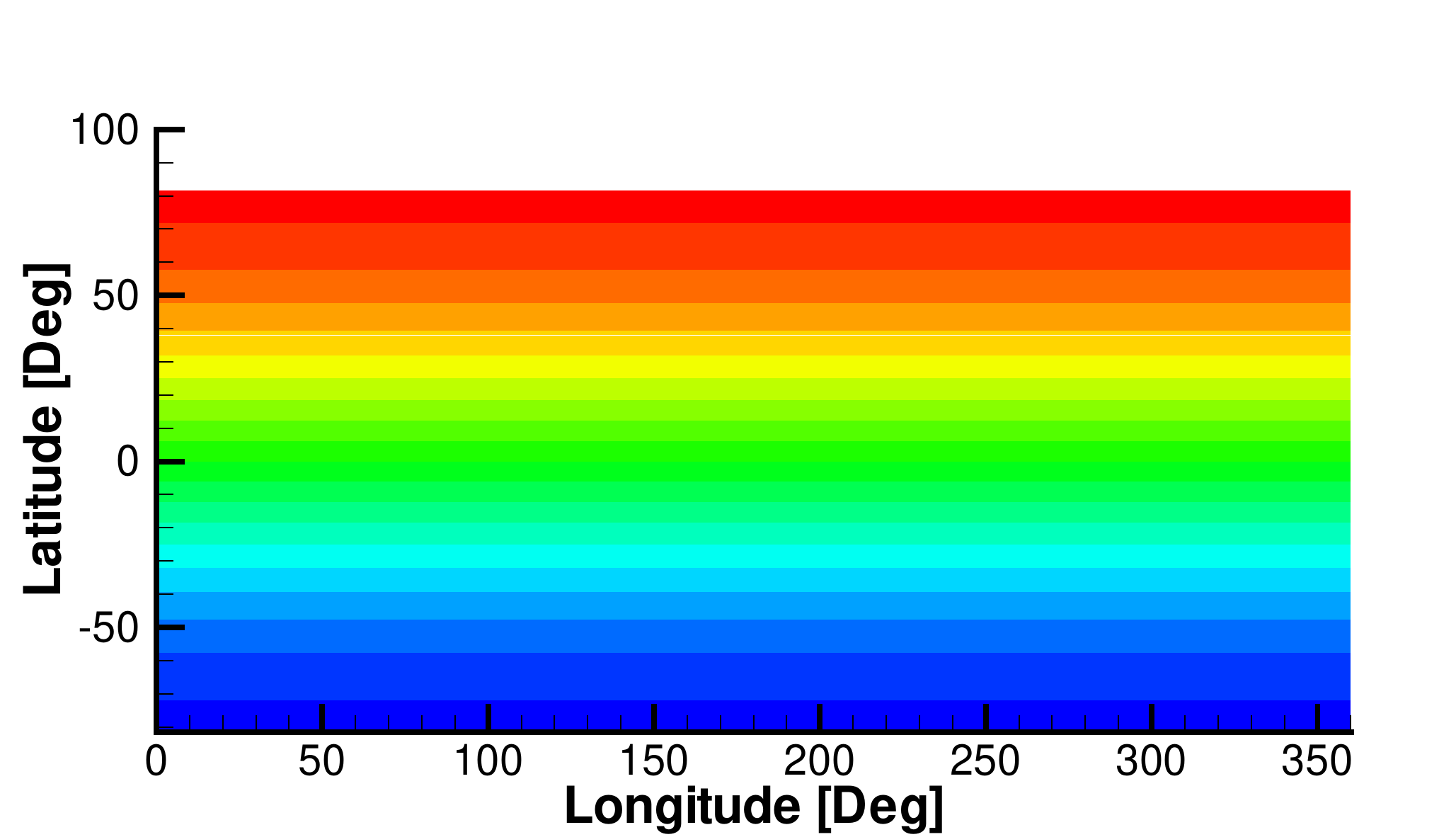} 
\includegraphics[trim = 0.3in 0.3in
  0.5in 3.in,clip, width = 2.12in]{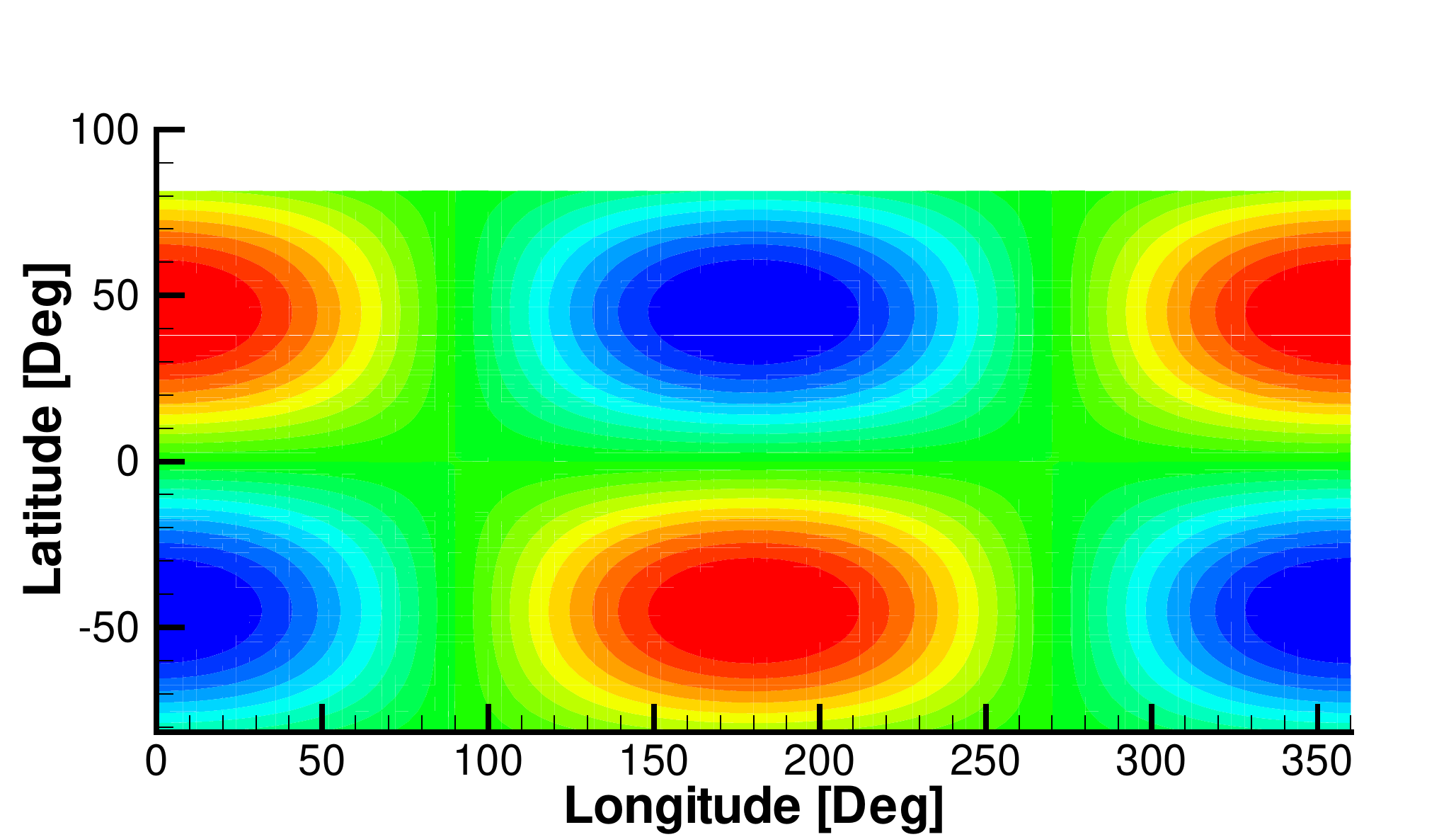} 
\includegraphics[trim = 0.3in 0.27in
  0.1in 3.in,clip, width = 2.12in]{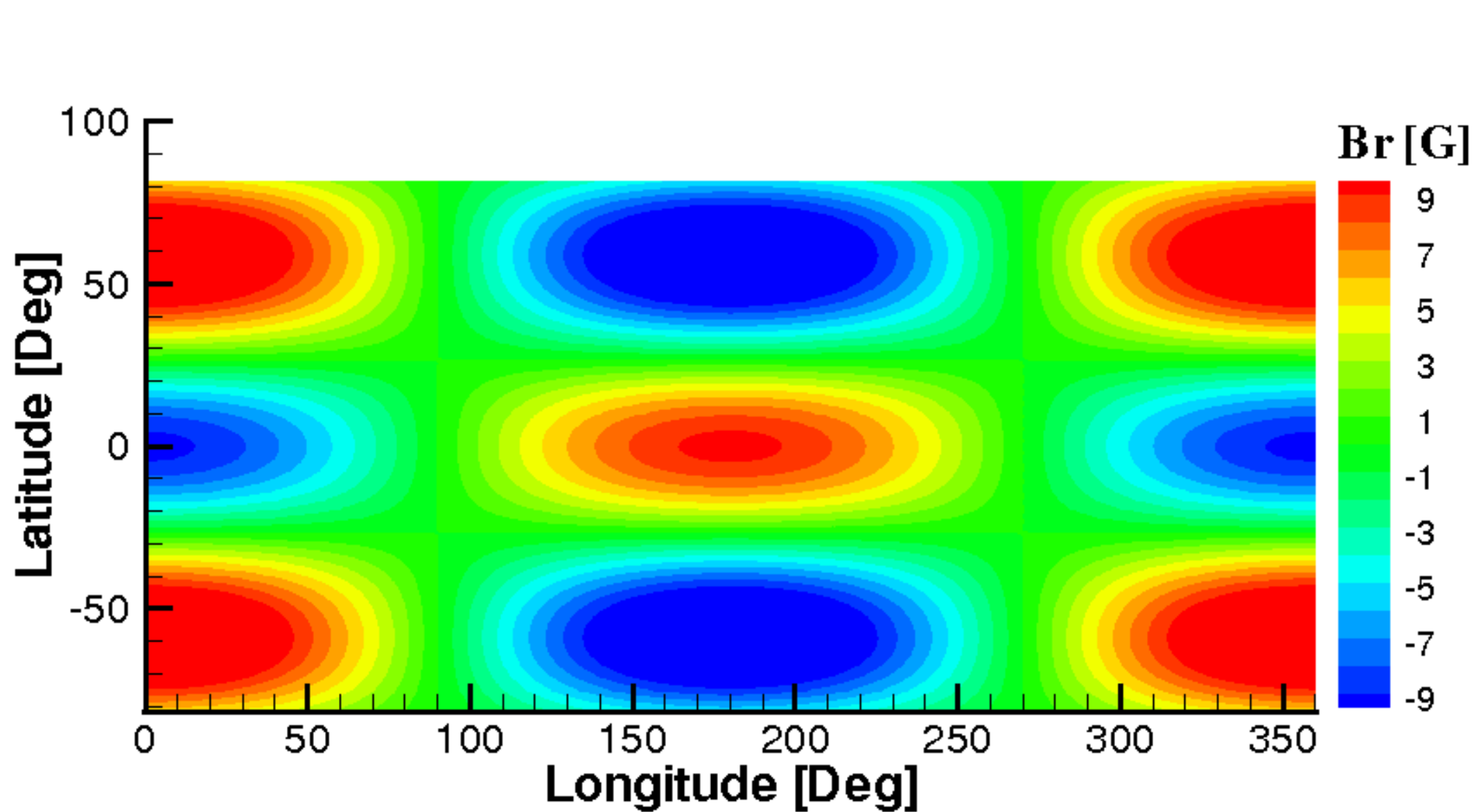} \\
\includegraphics[trim = 0.3in 0.3in
  0.5in 3.in,clip, width = 2.12in]{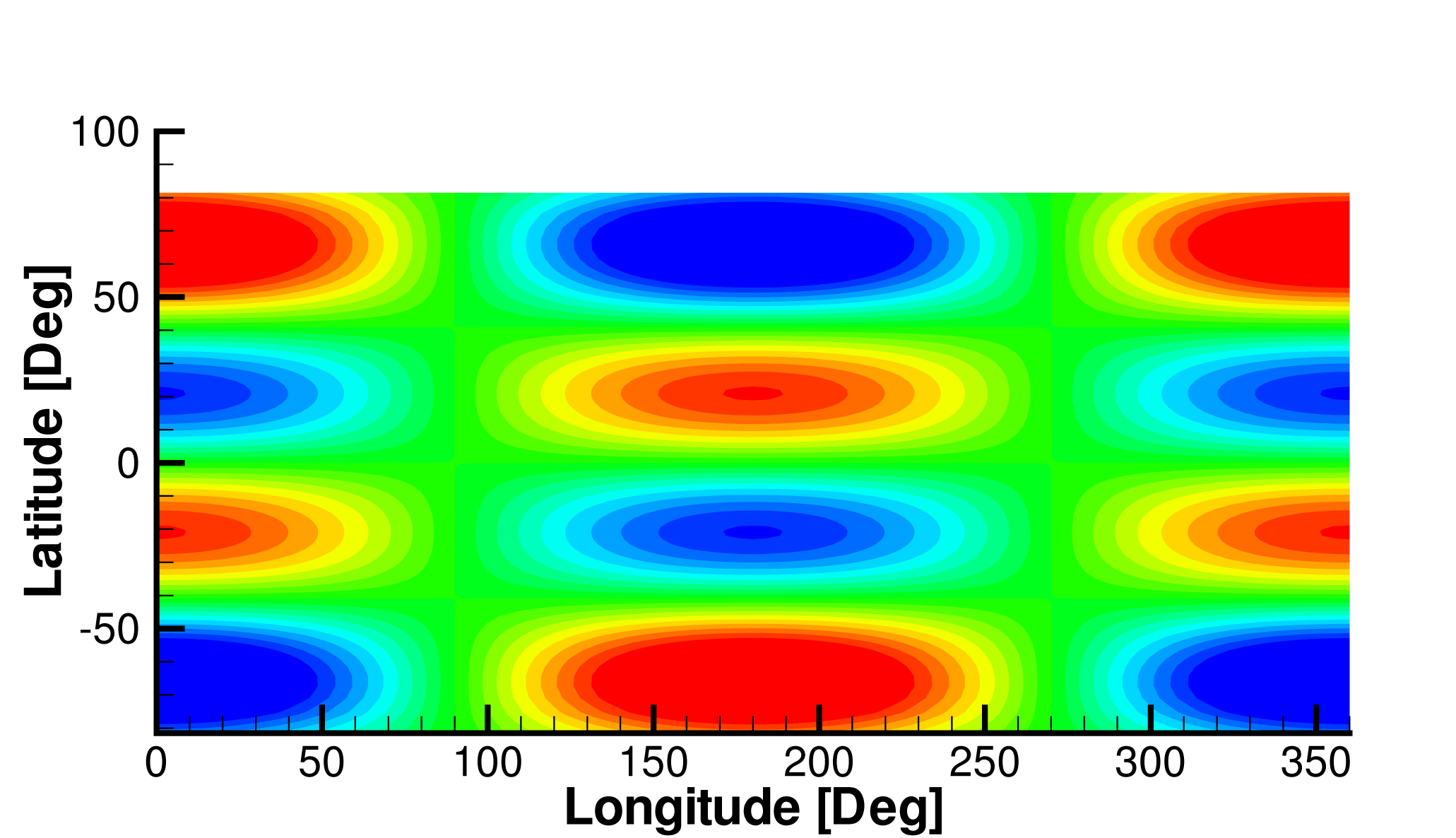} 
\includegraphics[trim = 0.3in 0.3in
  0.5in 3.in,clip, width = 2.12in]{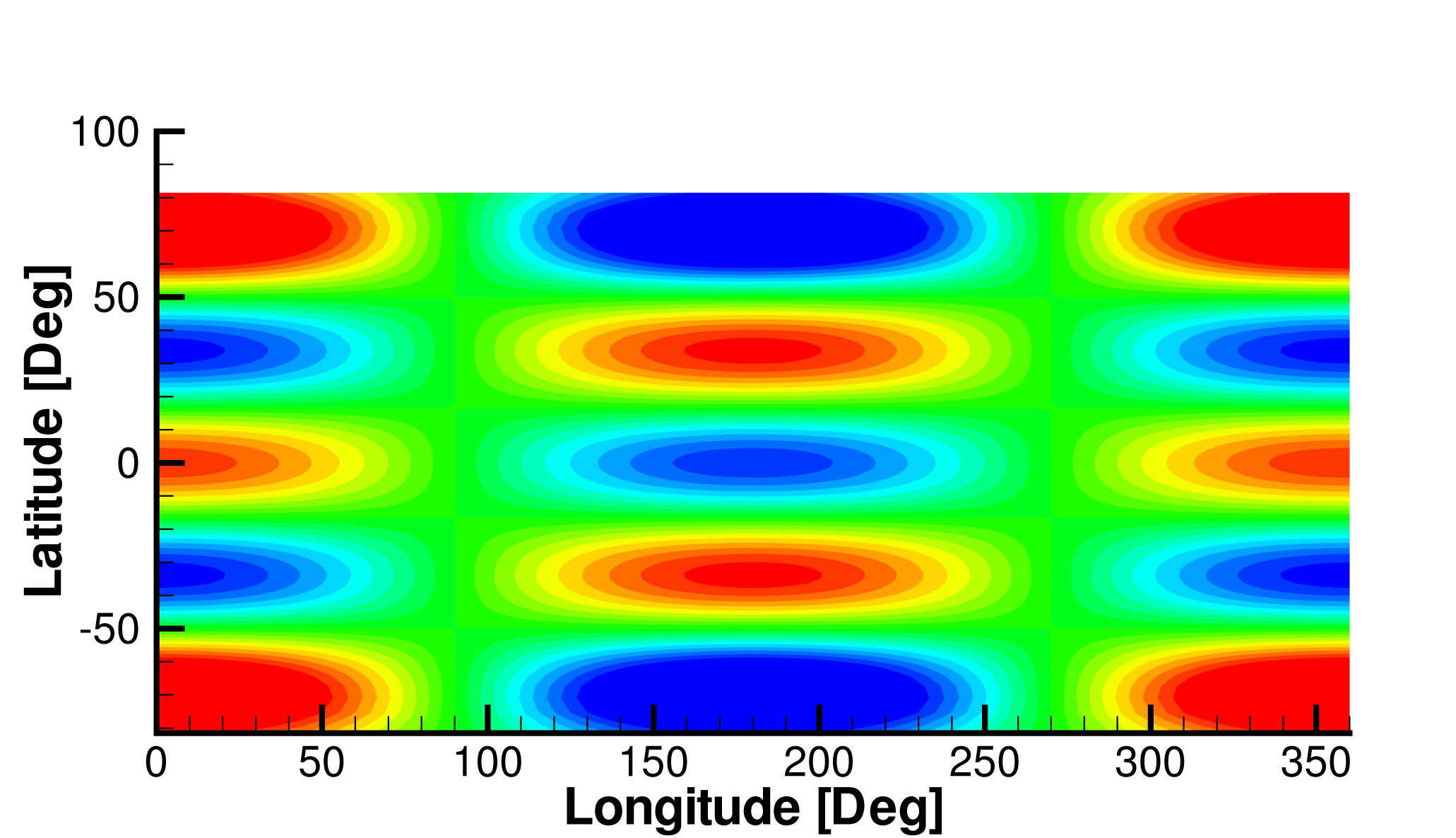} 
\includegraphics[trim = 0.3in 0.3in
  0.1in 3.in,clip, width = 2.12in]{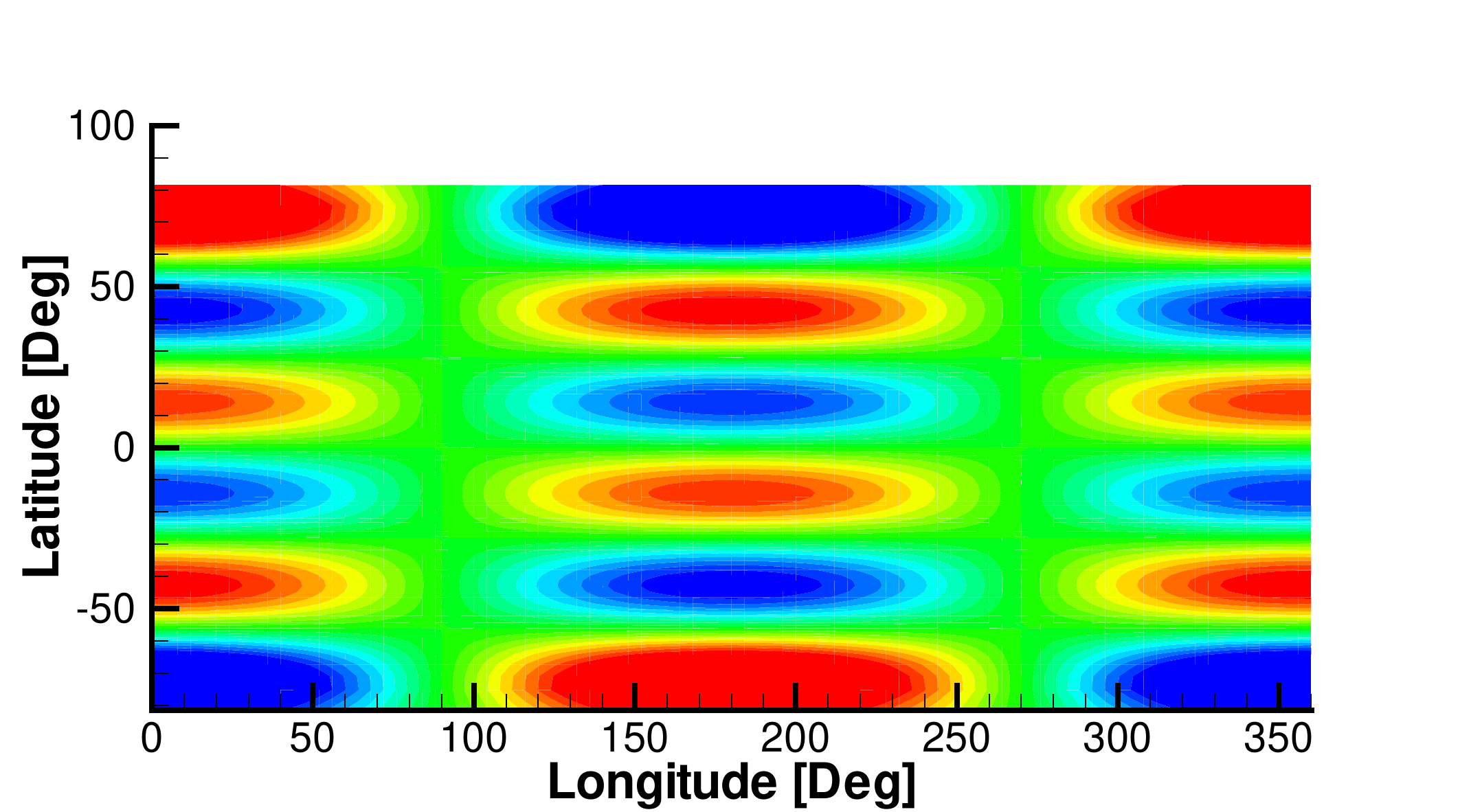} \\
\caption{Illustration of representative fiducial magnetograms for magnetic flux densities of increasing magnetic multipole
  orders (from top left to right bottom) up to order 6 for the 10~G
  amplitude case. Those for 20~G and
  100~G cases have the same appearance with rescaled amplitude.}
\label{fig:magnetograms}
\end{figure}

We perform three-dimensional stellar wind simulations for a hypothetical solar-mass star 
with a solar rotation period ($\sim 25$ days) and for different magnetic topologies and field
strengths.  Our grid of fiducial magnetograms consists of the ten first magnetic
 moments for peak field flux densities of $B=10$~G, 20~G and 100~G.  The maps for different
 morphologies are built using the corresponding term in the
 multipolar expansion, i.e., a Legendre Polynomial times a phase.   Examples of these fiducial magnetic maps
are shown in Figure~\ref{fig:magnetograms}.  
For a constant peak field strength, the integrated magnetic flux declines 
slightly with increasing magnetic moment, $n$, according to the orthogonality property
$ \int_{-1}^{1}P_m(x)P_n(x) dx = \frac{2}{2n+1} \delta_{mn}$.  
For the 20G baseline case, we therefore also compute magnetograms
normalized to the dipolar magnetic flux by the factor
$\sqrt{2/(2\cdot 1+1)}/\sqrt{2/(2n+1)}=\sqrt{(2n+1)/3}$. 
 
From the three-dimensional model solutions we extract the wind
density, $\rho$, and speed, $\mathbf{u}$, over the
Alfv\'en surface and at the stellar surface.   The Alfv\'en surface itself is determined by finding the surface for which the wind speed reaches the local Alfv\'en speed, $v_A=B/\sqrt{\rho}$,
neglecting the contribution of the electrons to the mass density, $\rho$, and for a pure hydrogen wind. 
We then compute
the mass and angular momentum loss rates at each point of the Alfv\'en
Surface, which in steady state, where all the gradients in
  pressure vanish, read:

\begin{equation}
 \frac{dM}{dt} = \rho \, (\mathbf{u} \cdot \mathbf{dA})
\end{equation}

\begin{equation}
\frac{dJ}{dt} = \Omega \, \rho \, R^2 \sin^2{\theta} \,(\mathbf{u} \cdot \mathbf{dA}),
\end{equation}
where $\frac{dJ}{dt}$ is the component of the angular momentum change in the
direction of the rotation axis, and is the only one contributing to a
change in the magnitude of $J$.  Here, $\mathbf{dA}$
refers to the surface element on the Alfv\'en Surface.
The formalism of \citet[][see also the form used by \cite{Vidotto14b}, Eqn (A6)]{Mestel99} includes an additional angular momentum term that
accounts for net stresses in the system, which in steady state become zero.

We also calculate the amount of open magnetic flux through
a spherical surface outside of the Alfv\'en surface.

\section{RESULTS}
\label{sec:Results}

\begin{figure*}
\center
\includegraphics[width = 4.4in]{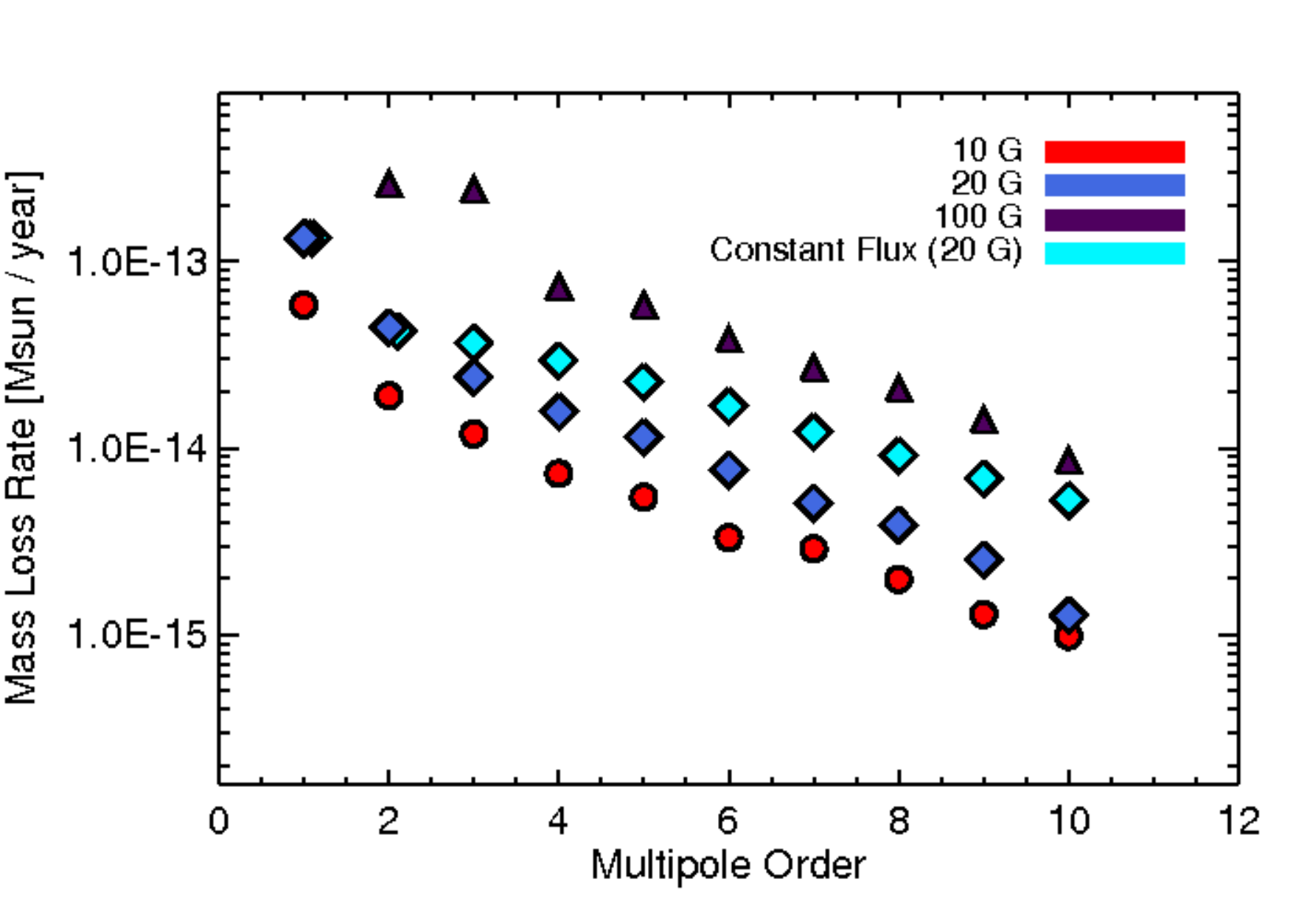}\\
\includegraphics[width = 4.4in]{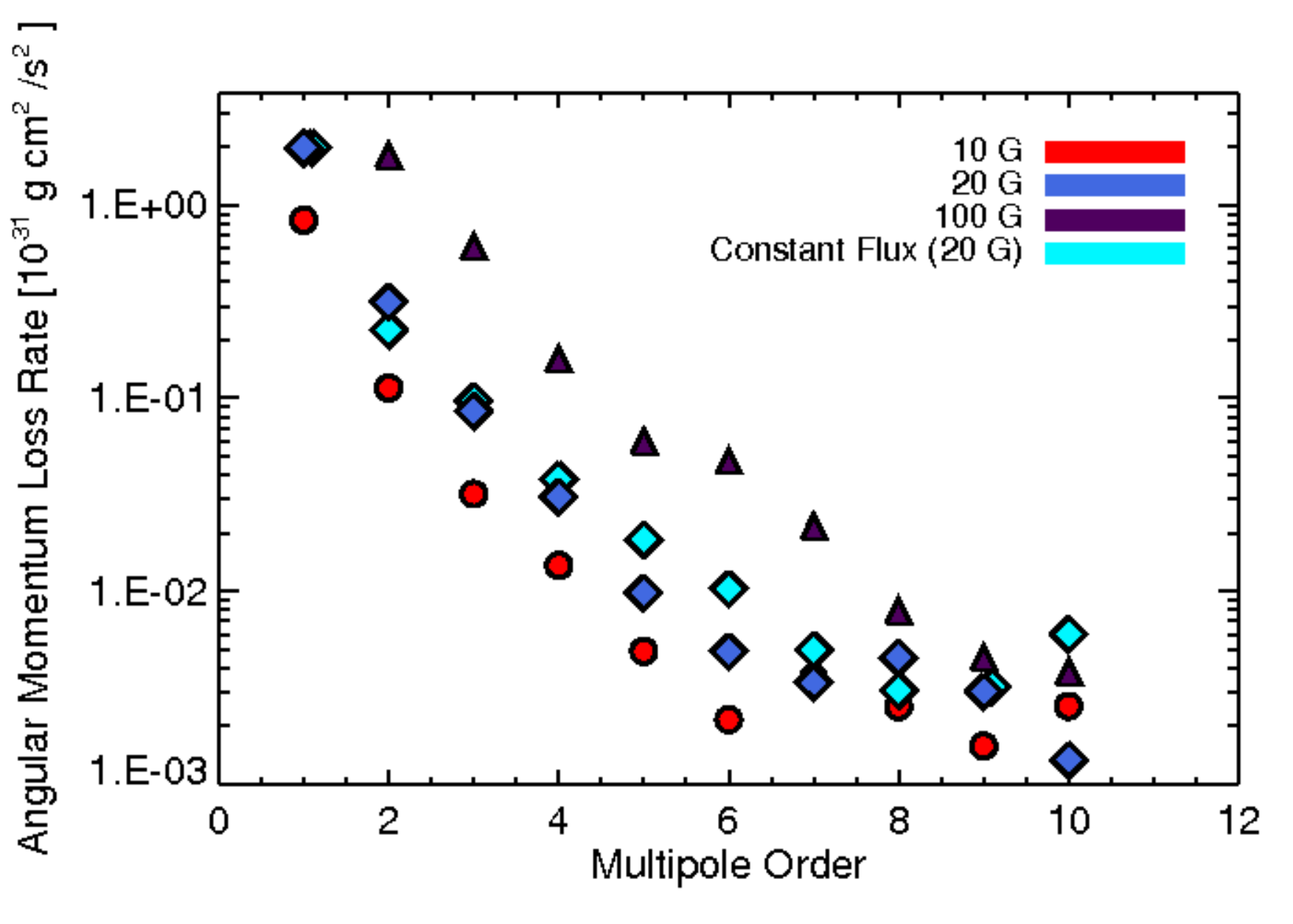}\\
\includegraphics[width = 4.4in]{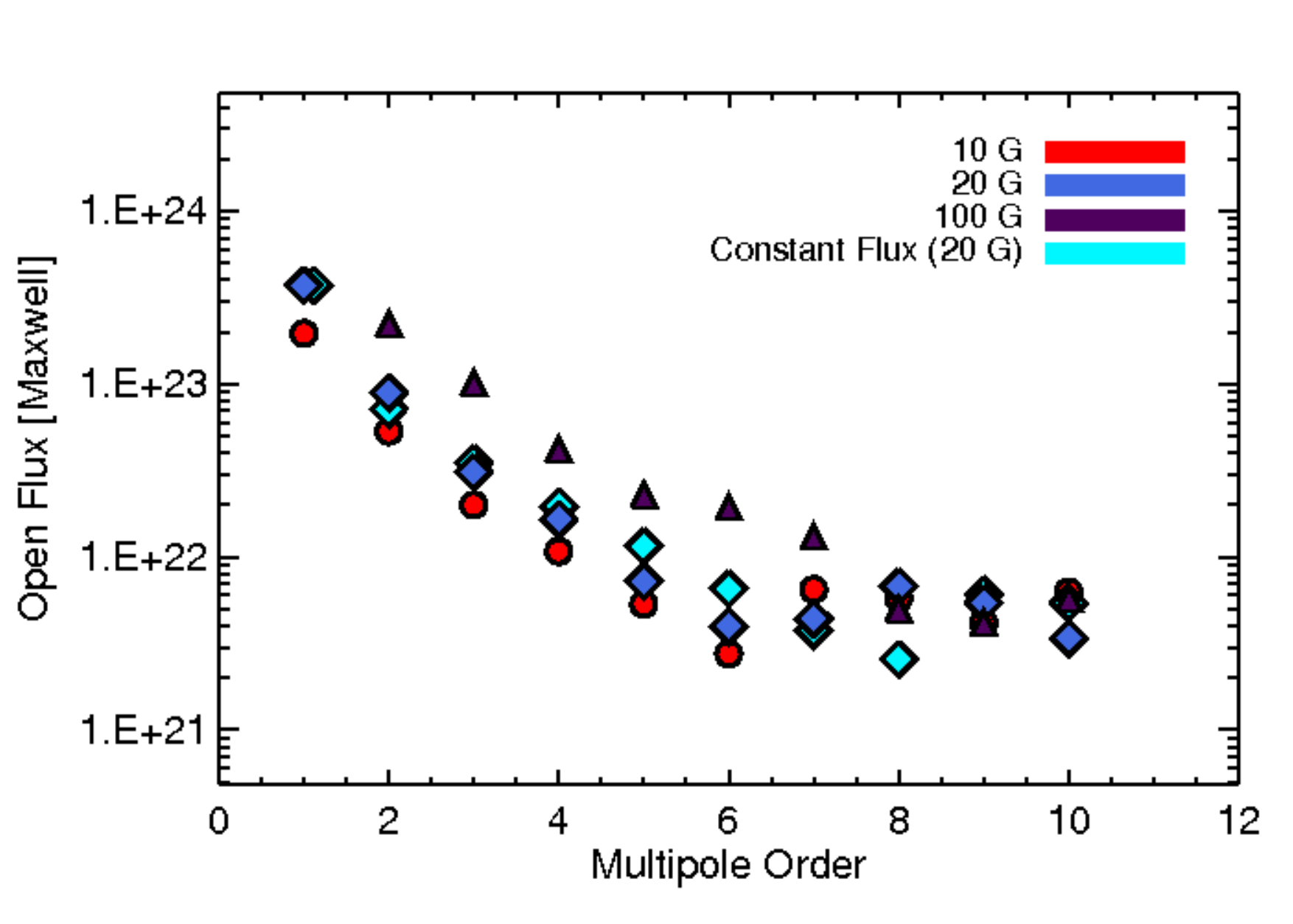}
\caption{Mass (top), angular momentum (middle) loss
  rates, and open flux (bottom panel) for different
  morphologies of 10~G (red), 20~G
  (blue) and 100~G (purple) field strengths.  Cyan points correspond to a constant magnetic flux normalized to that of a dipolar 20~G field strength.}
\label{fig:aml}
\end{figure*}

The mass and angular momentum loss rates, together with 
amount of open flux computed for each case in our grid of models, are plotted in Figure~\ref{fig:aml}.

The most dramatic result is a
systematic decrease of both $\dot{M}$ and $\dot{J}$ by two and
three orders of magnitude, respectively, with increasing complexity of the magnetic topology. 
For a given magnetic field strength, the logarithmic mass loss rate scales approximately linearly with magnetic moment, $\log \dot{M}\propto n$.
Angular momentum loss is instead not a simple function of mass loss when
including higher order magnetic moments but is very well correlated
with the amount of open flux.  
The reasons underpinning these trends become clear when examining the details of the MHD model solutions and how the open field regions from which the wind is driven change with magnetic moment.   The trends with constant magnetic flux are slightly less steep than for constant peak field, as expected from the general dependence of $\dot{M}$ and $\dot{J}$ on field strength for a given order of complexity. 

\begin{figure*}
\center
\includegraphics[trim = 0.2in .2in
  0.5in 0.2in,clip, width = 3.2in]{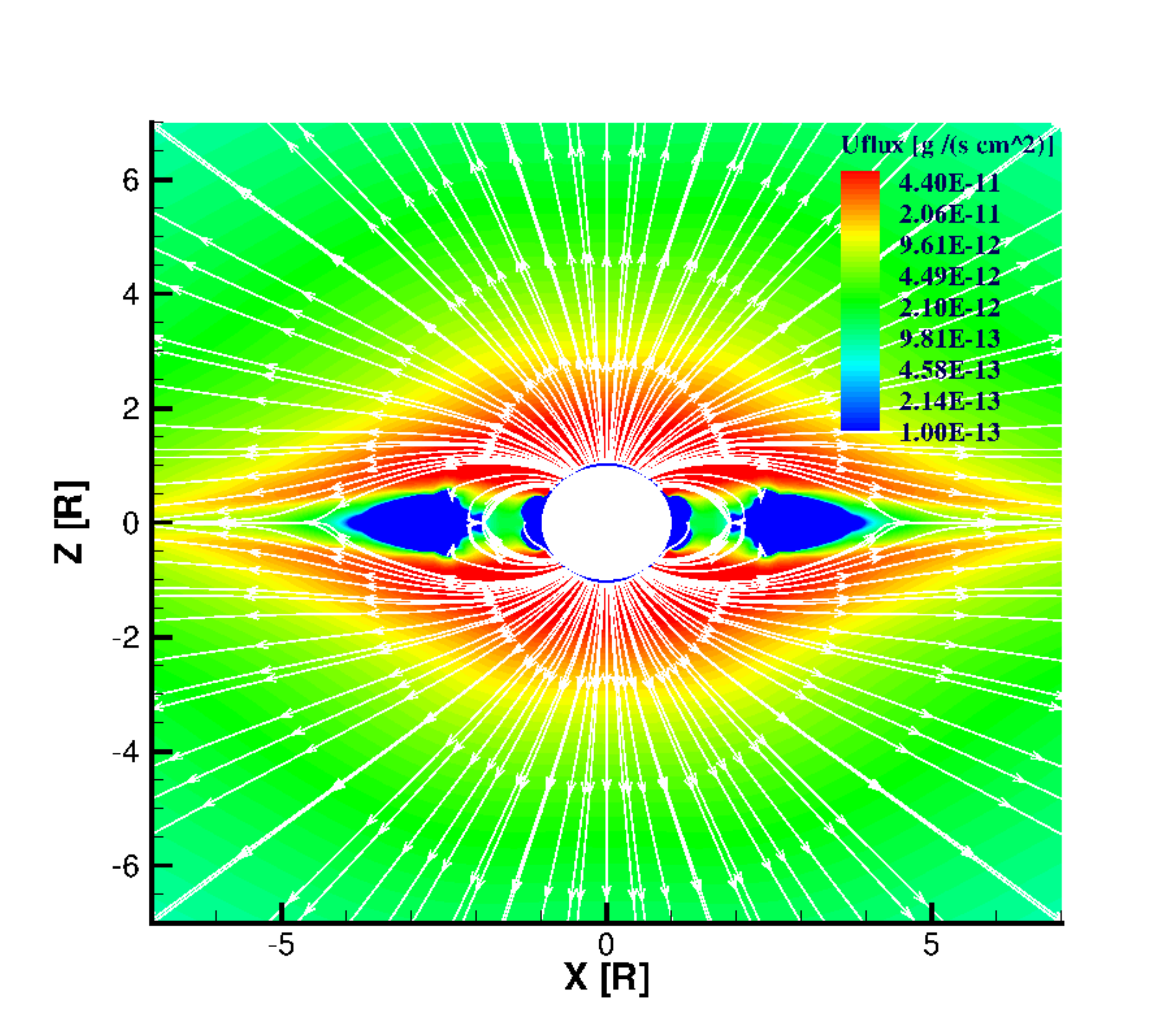} 
\includegraphics[trim = .2in .2in 
0.5in 0.5in,clip,width =3.2in]{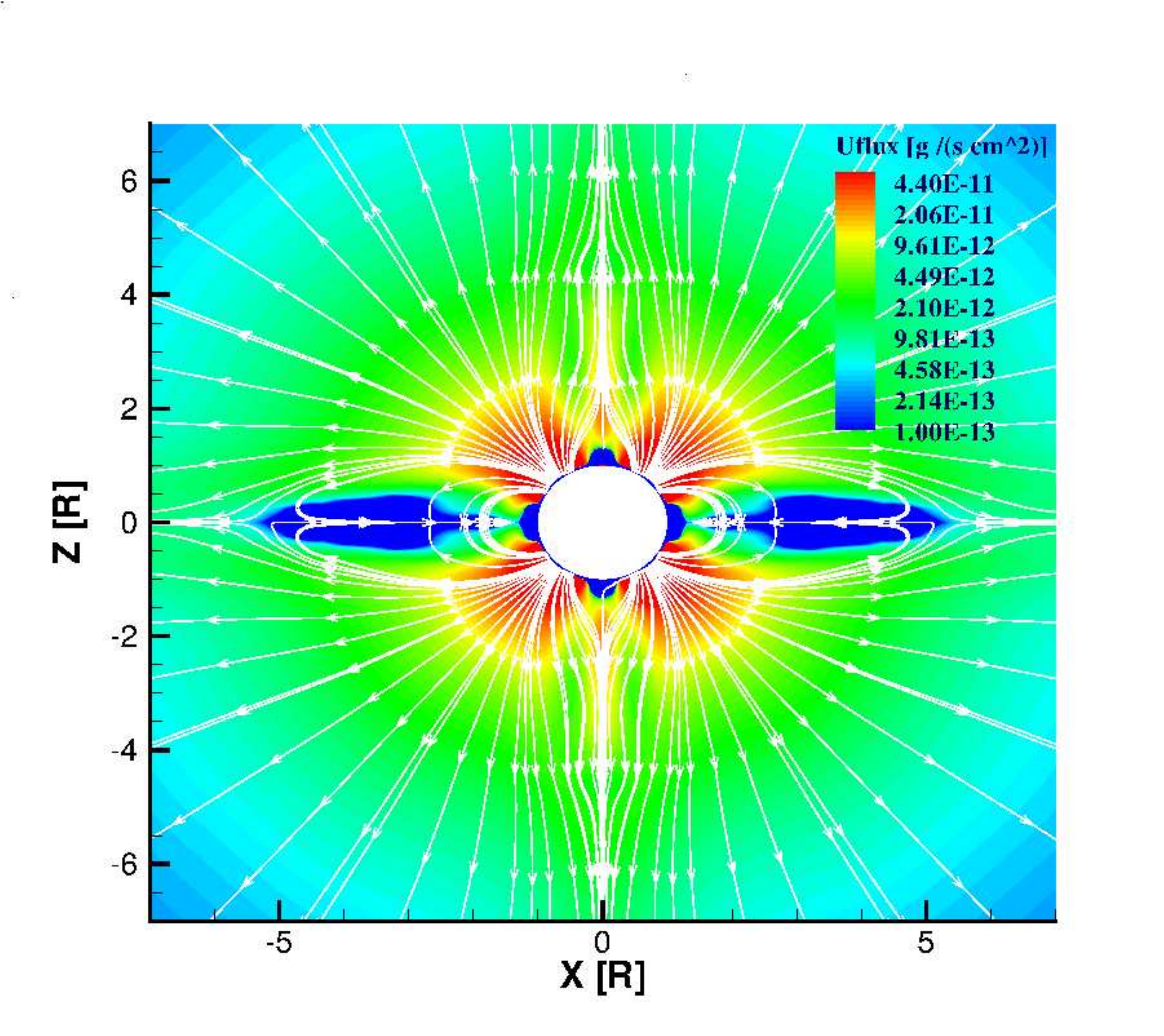}\\
\includegraphics[trim = .2in .2in
  0.5in 0.5in,clip, width = 3.2in]{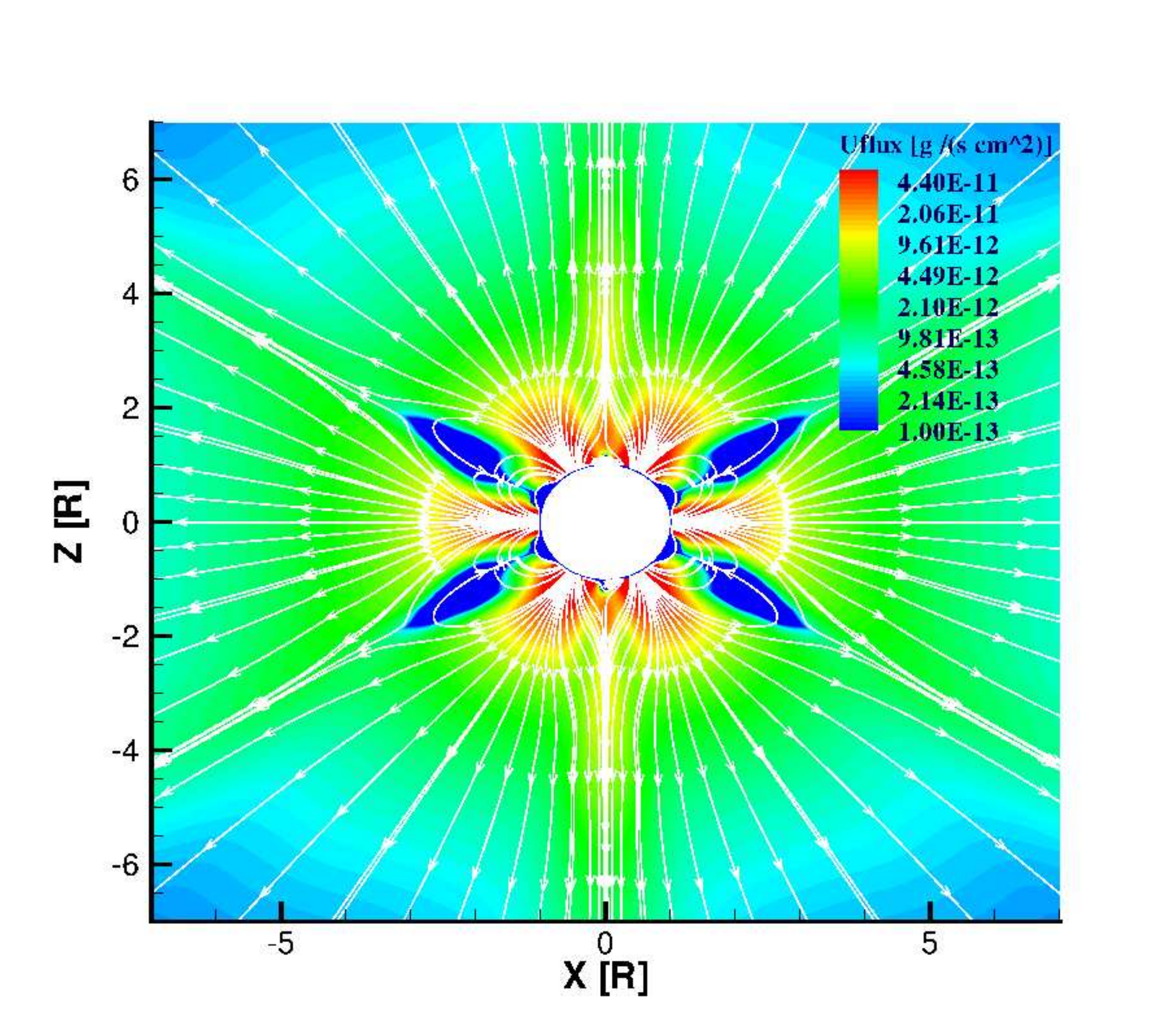} 
\includegraphics[trim = .2in .2in
0.5in .5in,clip,width =3.2in]{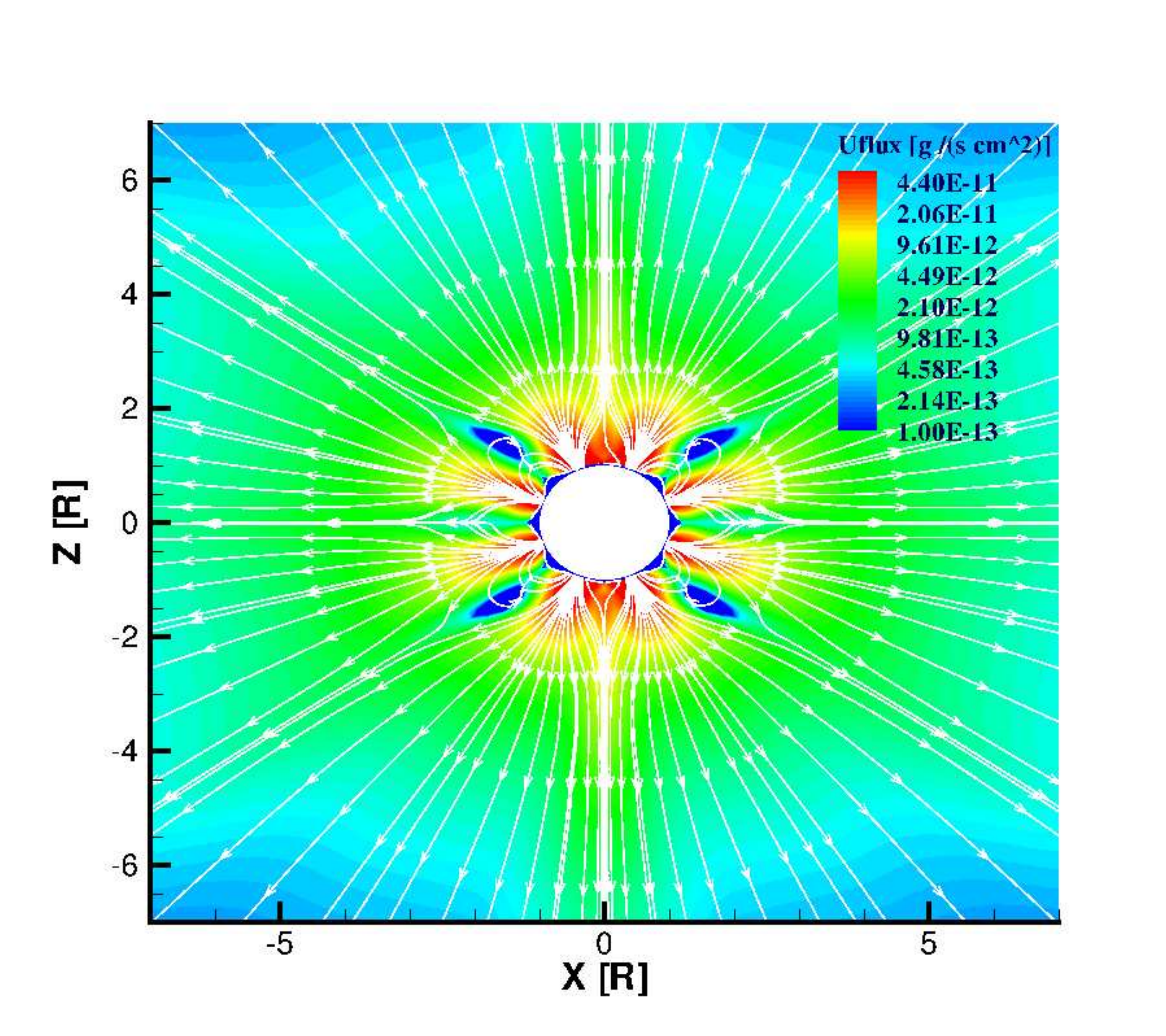}\\
\includegraphics[trim = .2in .2in
0.5in 0.5in,clip,width =3.2in]{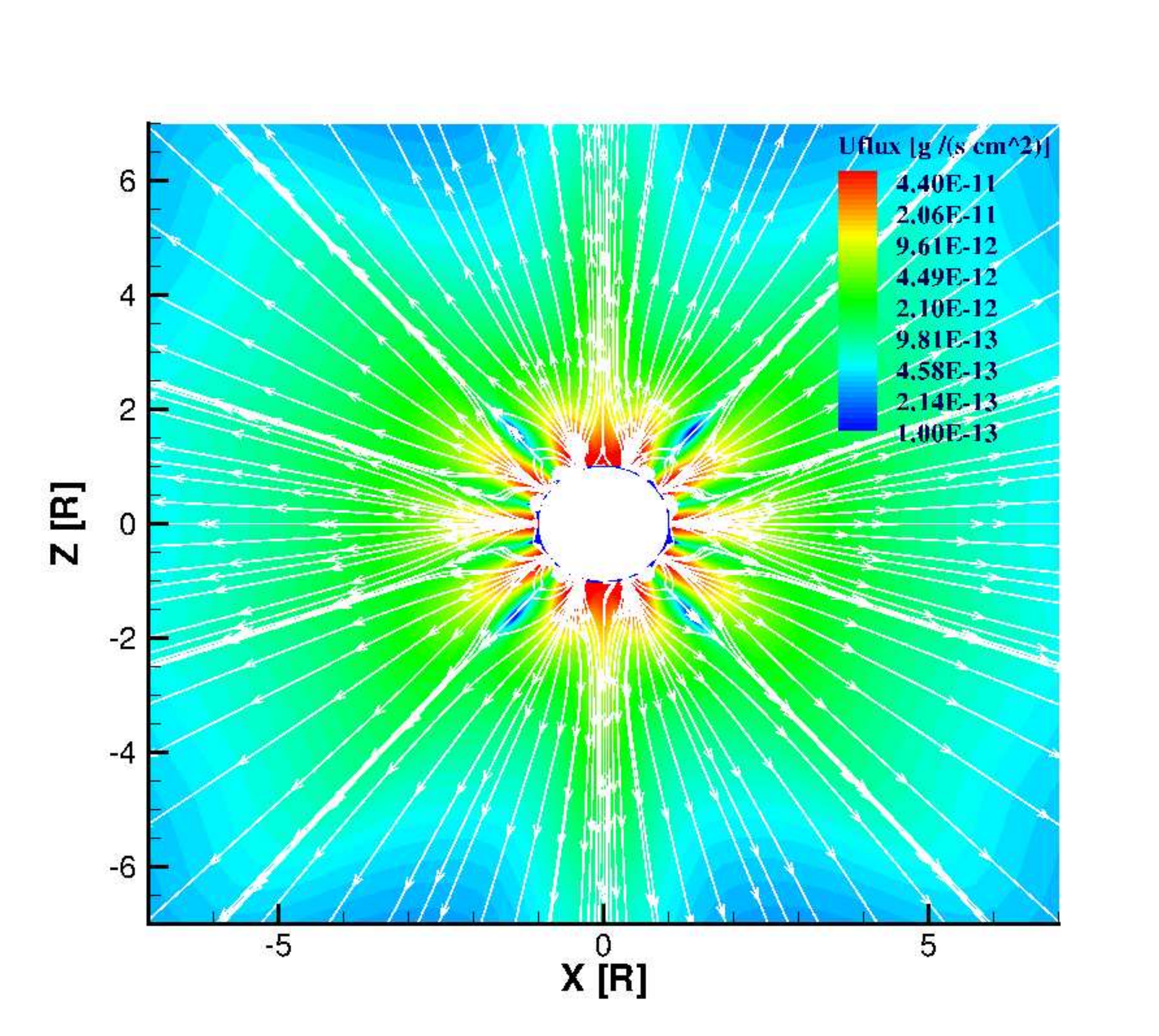}
\includegraphics[trim = .2in .2in
0.5in 0.5in,clip,width =3.2in]{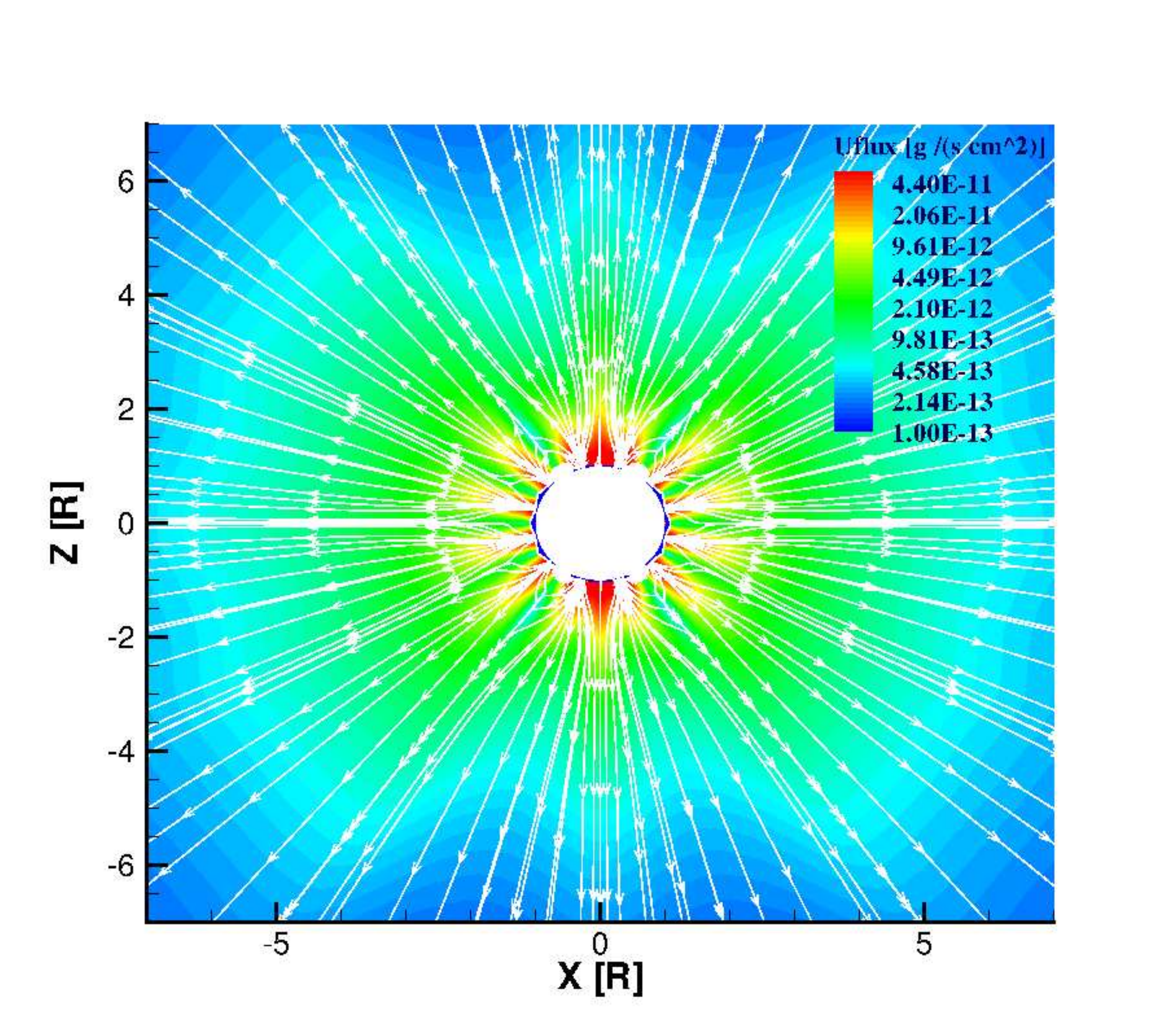}\\
\caption{Meridional cut of the wind mass flux for increasing magnetic
  complexity for 20~G models, from a
  dipole (top left) to a 6th order multipole (bottom right).
  Magnetic field lines are plotted in white and axis
  units are in stellar radii.}
\label{fig:WS}
\end{figure*}

\begin{figure*}
\center
\includegraphics[trim =0.1in 0.1in 
0.1in 0.1in,clip,width =4.5in]{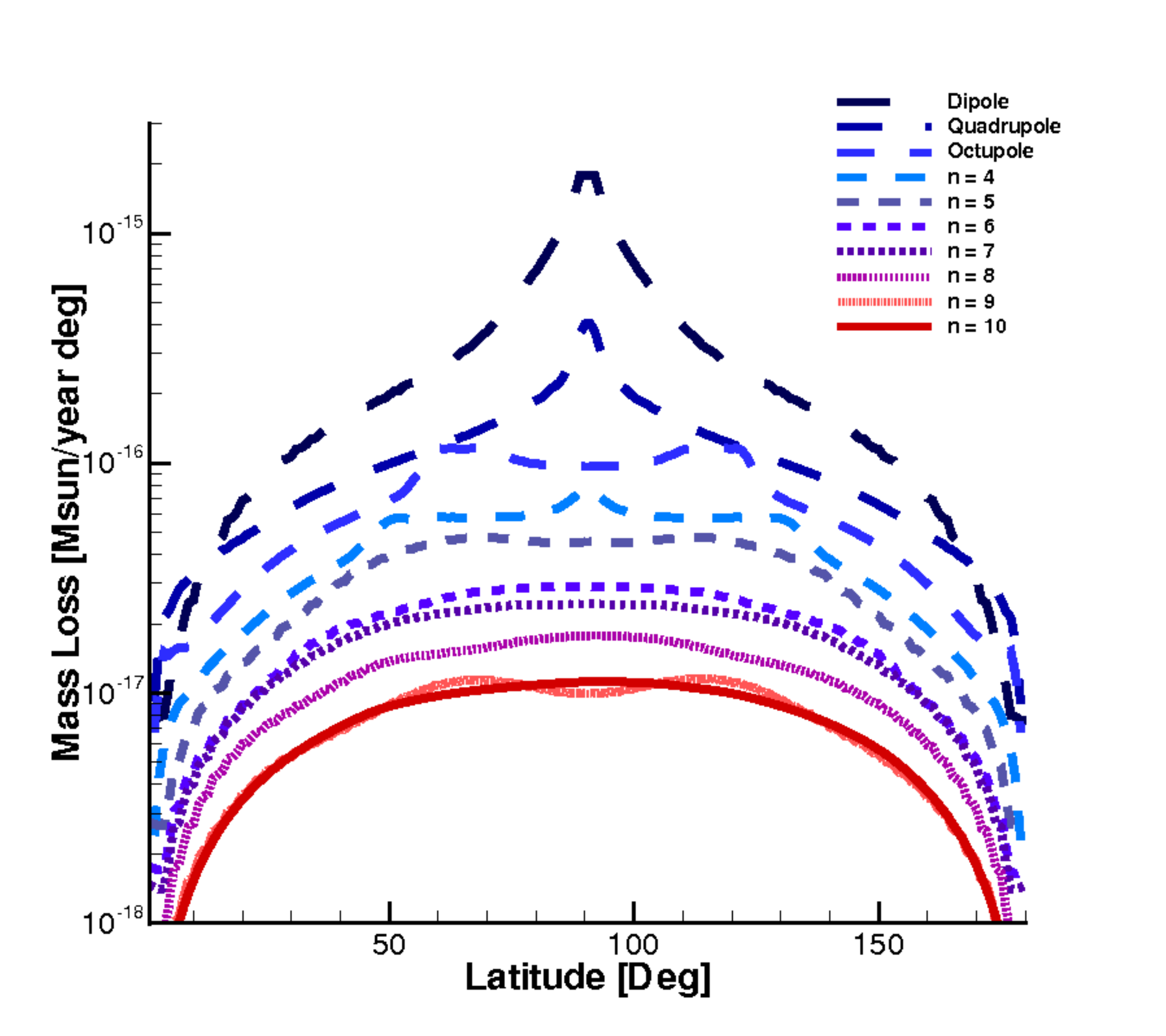}
\caption{Mass loss distribution in latitude for different morphologies of 20G magnetic field strength, where $n$ refers to magnetic multipole order.}
\label{fig:Mlat}
\end{figure*}

There are three interrelated aspects to the angular momentum
loss: the mass flux, the Alfv\'en radius over which it acts as a rotational brake, and the
latitude at which the mass release happens.  
In order to compare the sizes and latitudes of the closed field line
regions for the different magnetic morphologies, Figure~\ref{fig:WS}
shows meridional cuts of the wind mass
flux together with selected magnetic field lines for the 20~G models. 
The area of the stellar surface occupied by open field decreases dramatically with increasing
magnetic complexity, while open field regions ( "coronal
holes") also get distributed more homogeneously over latitude.  
Only the first affects $\dot{M}$ directly, but both are expected to lead to a reduction of $\dot{J}$. The open field lines spread and overlay regions of closed field and the mass flux at the apex of the closed loop systems is larger than in the middle of open field regions.  For the dipolar case this happens only at the
equator, resulting in an equatorially-dominated mass loss,
which is the most efficient latitude for losing angular momentum.  This is also seen in Figure~\ref{fig:Mlat}, where mass loss
is plotted as a function of latitude for the 20~G simulations.  Increasing magnetic moment leads to a larger number of less pronounced local maxima.  This change in latitudinal dependence of the mass loss leads to a stronger dependence of angular momentum loss on the magnetic complexity than just the mass loss rate itself (see Figure~\ref{fig:aml}).  
 
\begin{figure*}
\center
\includegraphics[trim = 1.9in 1.in
  2.7in 2in,clip, width = 2in]{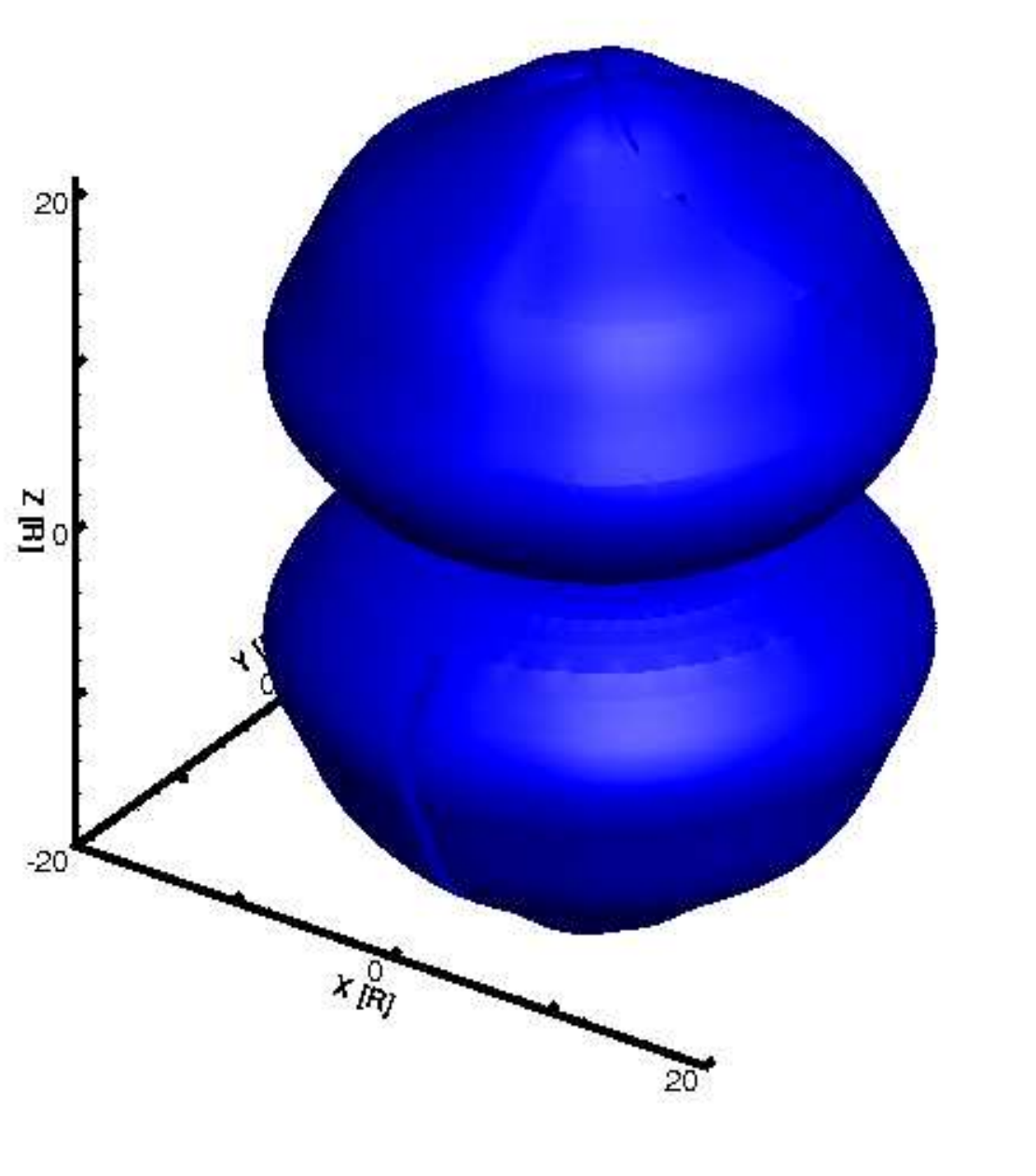} 
\includegraphics[trim = 1.9in 1.in 
2.7in 2in,clip,width =2in]{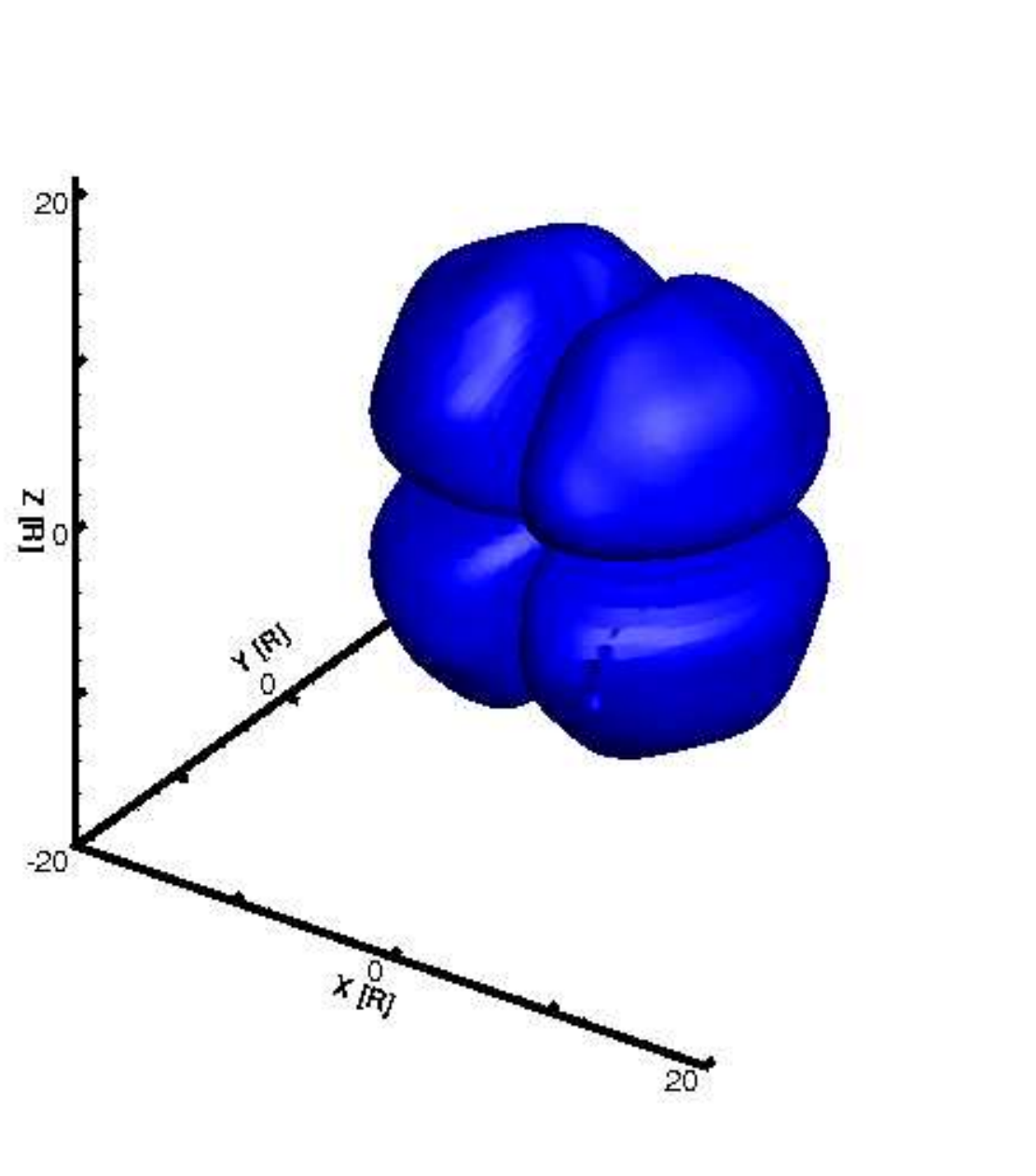}
\includegraphics[trim = 1.9in 1.in
  2.7in 2in,clip, width = 2in]{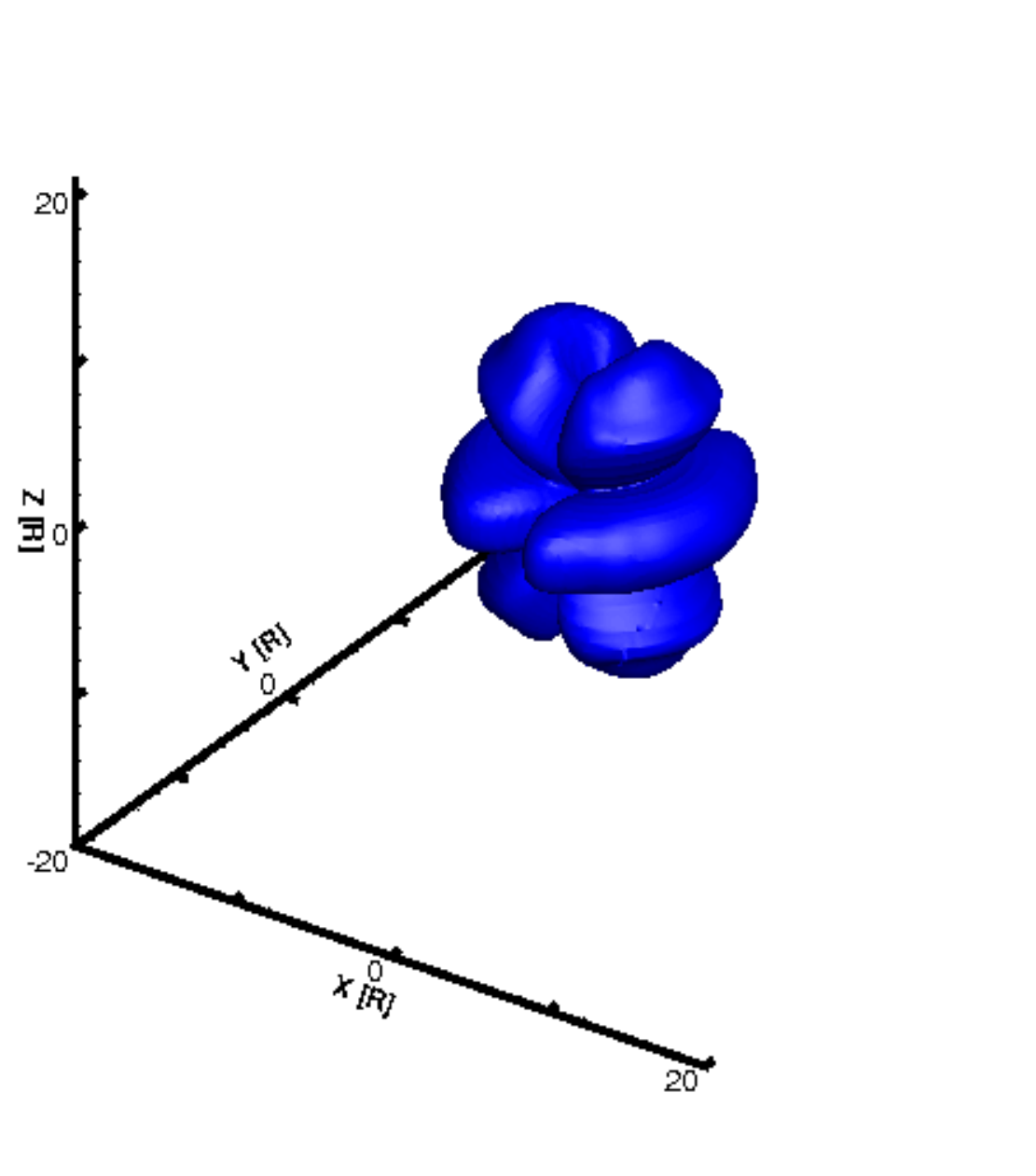}\\ 
\includegraphics[trim = 1.9in 1.in
2.7in 2in,clip,width =2in]{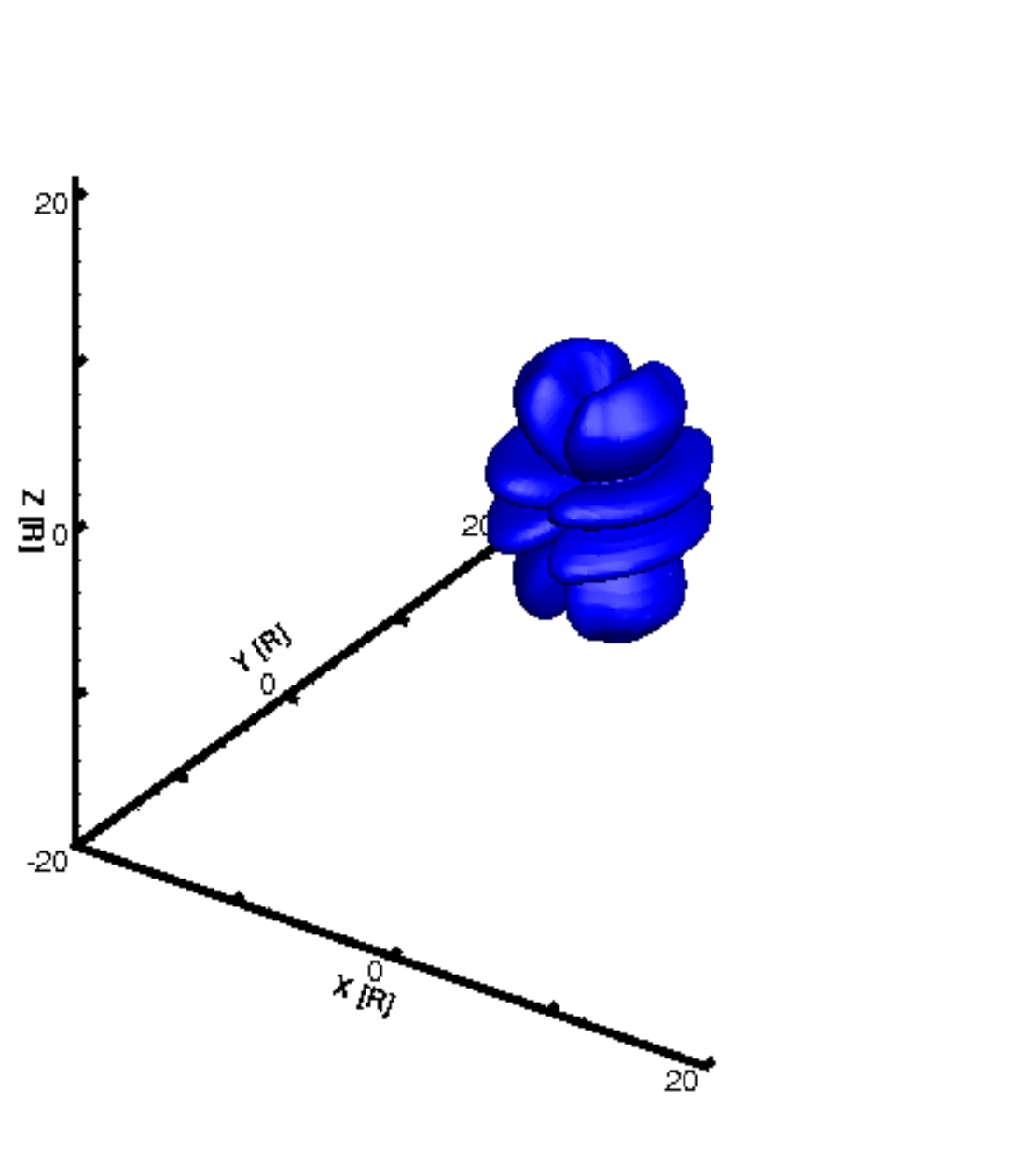}
\includegraphics[trim = 1.9in 1.in
2.7in 2in,clip,width =2in]{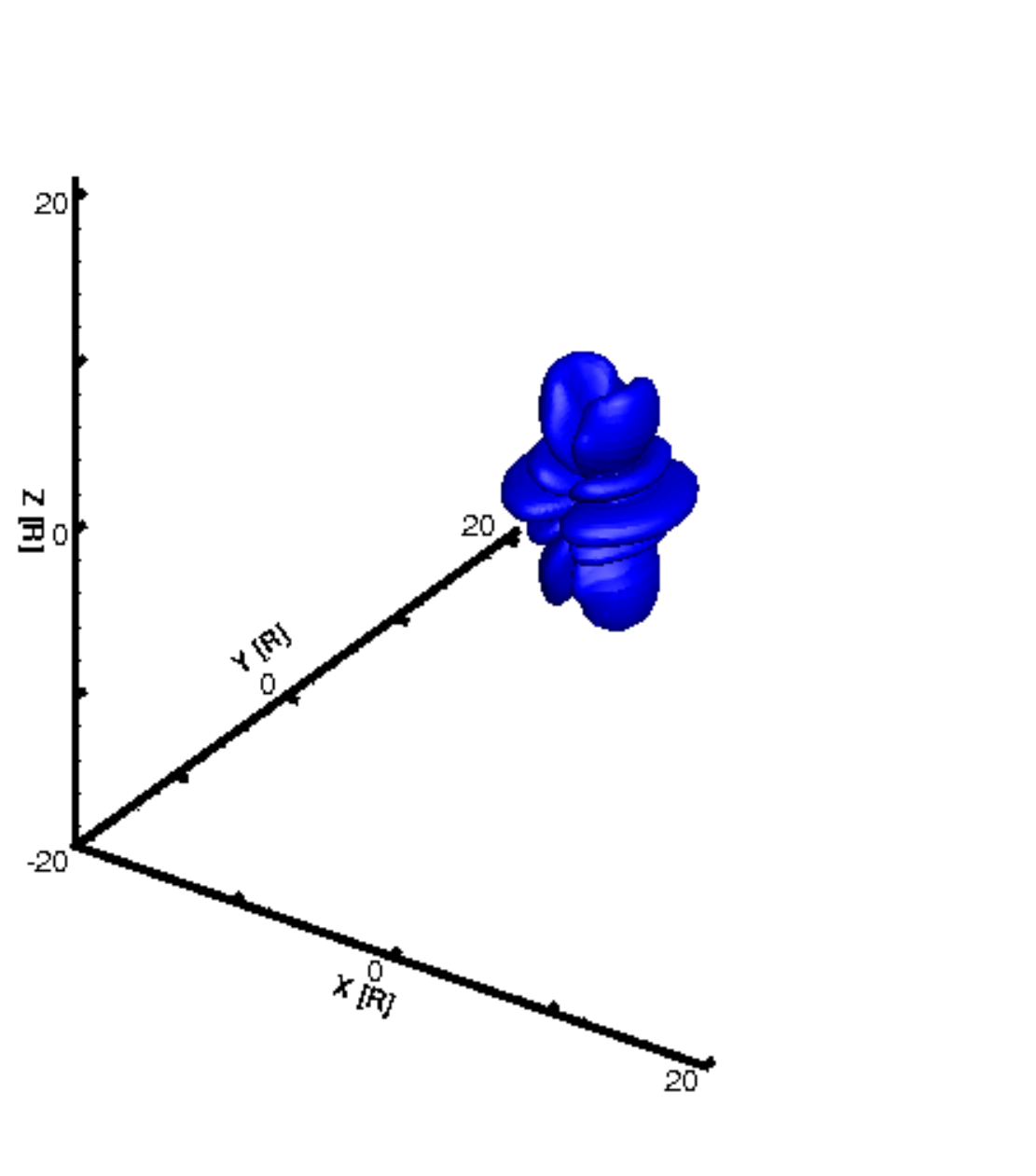}
\includegraphics[trim = 1.9in 1.in
2.7in 2in,clip,width = 2in]{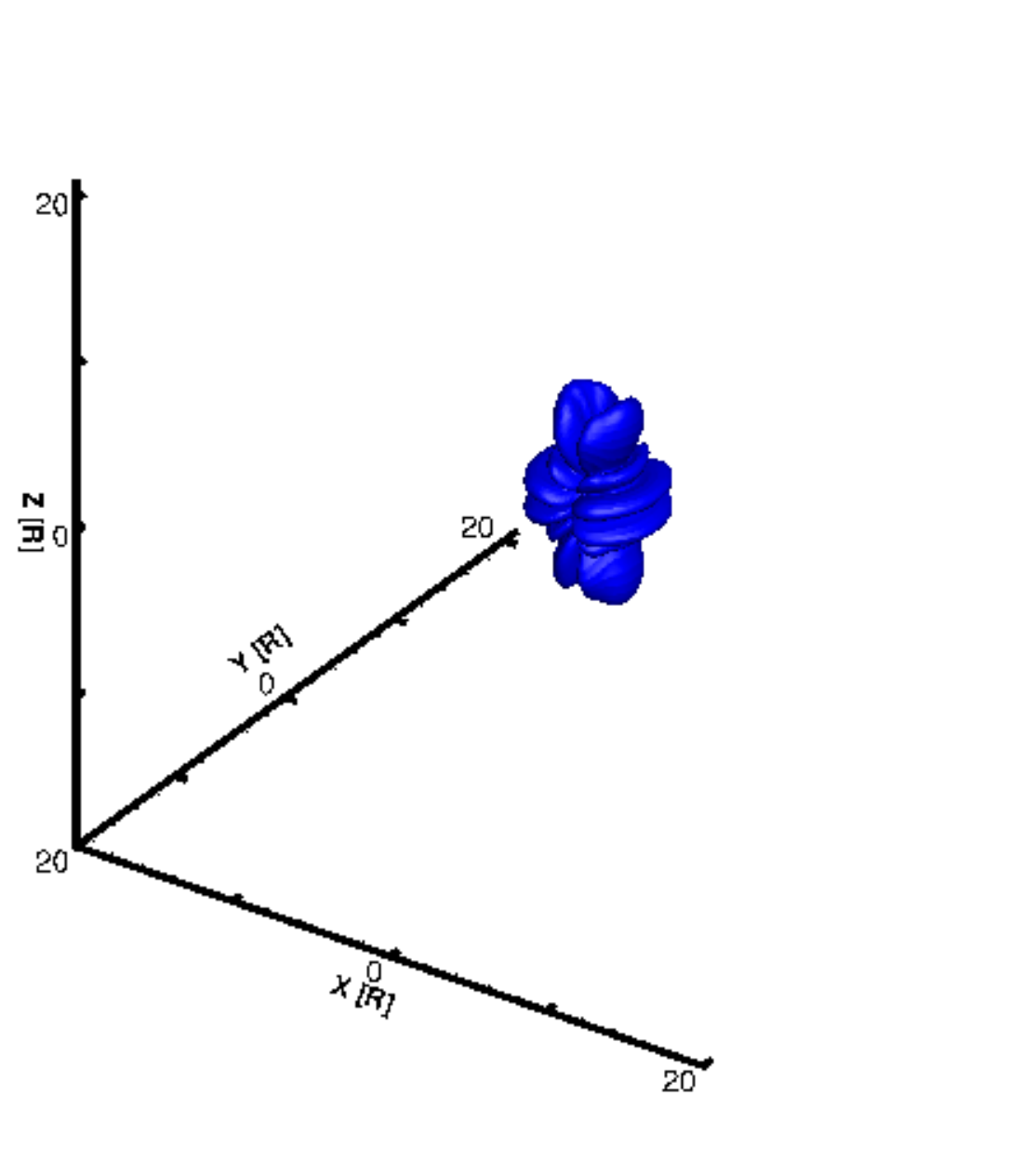}
\caption{Three dimensional Alfv\'en surfaces for 20~G fields of increasing 
 complexity, from a dipole (top left) to
  a 6th order magnetic multipole (bottom right).}
\label{fig:AS}
\end{figure*}

Solar wind observations show that the fast and less dense wind originates from the center of coronal holes, while the slow and denser solar wind originates from the boundary between the coronal holes and the closed loops \citep[e.g.][]{Phillips95,McComas07}.  Our results show that the size of the coronal holes decreases with the increase of the multipole order. As a result, the fast wind in the solution is eliminated to a point where only slow wind exists. 
On the other hand, the radial dependence of the magnetic field is proportional to
$1/r^{n+1}$, where $n$ is the magnetic multipole order.  Therefore, the
magnetic flux falls faster with radius for higher orders.  As a
consequence one should expect the Alfv\'en velocity at a given radial distance to decrease with
magnetic complexity (much more than the wind speed), and the Alfv\'en radius to become
smaller.  Figure~\ref{fig:AS} shows how the three dimensional Alfv\'en surface of
different magnetic morphologies (for the 20~G models) shrinks rapidly with increasing 
complexity.  The reduction in the wind speeds translates to 
a reduction of $\dot{M}$ and, therefore, $\dot{J}$,  while
the reduction in the lever arm due to the shrinkage of the Alfv\'en
surface amplifies the reduction of $\dot{J}$ with higher magnetic multipole orders.


\section{DISCUSSION}
\label{sec:Discussion}

\citet{Brown14} has discussed in detail the difficulties existing stellar spin down models have in matching the observed distributions of stellar rotation velocities in young open clusters.  Core-envelope decoupling models are able to provide a reasonable match with specified percentile points in the distribution of rotation periods, $P_{rot}$, but fail in reproducing  its bimodal aspect and how it changes with time.   

\citet{Barnes03} dubbed the bimodal branches $C$ for the rapid rotators and $I$ for the slower ones, tentatively identifying these as being dominated by convective and interface dynamos, respectively.  He speculated that on the $C$ sequence, the magnetic coupling of the star to the wind was weak, possibly due to the magnetic field morphology.   \cite{Barnes.Kim:10} fleshed out the ideas of  \citet{Barnes03} and the core-envelope decoupling idea, synthesizing a purely descriptive expression for period evolution from the two different regimes that \citet{Brown14} referred to as the Symmetrical  Empirical Model (SEM).   While the SEM appears to be able to match the general distribution of periods seen in open clusters, \citet{Brown14} concludes that, as specified, it also fails to match the details of the bimodal morphology, and in particular in being able to sustain G stars with ages above 200 Myr and $P_{rot} \leq 2$~days, as are observed in M34 and M37. 

\citet{Brown14} emphasizes that existing models might be improved to better reproduce observations, but for simplicity introduced a different concept that amounts to an {\it ad hoc} mass-dependent transition probability between the $C$ and $I$ states.  The MDM, briefly introduced in Section~\ref{sec:Intro}, posits that stars are born
with their magnetic dynamos operating in a mode that couples very
weakly to the stellar wind and, at some point, this mode spontaneously and randomly changes to a strongly-coupled mode.
Stars then spin down following the torque law, 
\begin{equation}
\frac{dJ}{dt} = K_M \Omega^3 f^2(B-V),
\end{equation}
where, as in the SEM model \citep{Barnes.Kim:10}, $B-V$ stands for
color and $f^2(B-V)$ is a function that
depends on stellar parameters.
The constant $K_M$ may take two values corresponding to the strong and
the weak-coupling regimes, the latter smaller than the former by a factor of 100 or more.  One major difference between the MDM concept and other models is that the introduction of a random transition between the different modes means that the early evolution of $P_{rot}$ for a given star is not a simple deterministic function of its initial rotation rate.  While faring less well in matching the details of the $I$ sequence, the MDM model appears successful in qualitatively matching the bimodal structure of $P_{rot}$ distributions at young ages.  What the model lacks is a physical mechanism responsible for the coupling change. 

Our MHD simulations indicate that
the efficiency of angular momentum loss is strongly suppressed for
stars with complex surface magnetic fields.   The quadrupolar case already represents almost an order of magnitude decline compared with a dipolar field, irrespective of the absolute field strength, while for high order fields both mass and angular momentum losses are almost independent of the magnetic field strength.
The results have a conspicuous parallel to the coupling constant proposed by \citet{Brown14}, and indicate that, as has already been mentioned qualitatively in the past \citep[e.g.][]{Taam89, Barnes03,Vidotto14a}, magnetic complexity could be a key missing factor required to understand stellar rotation.  
In the light of growing 
evidence pointing to a systematic increase of higher-order
magnetic moments in younger stars, we propose that the evolution of magnetic morphology on the
stellar surface (determined by the evolution of the stellar dynamo)
provides a simple explanation for the quite different $\dot{J}$ 
regimes that appear to be required to match observations of early stellar rotation evolution.

If we identify $K_M$ with magnetic complexity, 
for a solar-like star with a constant $\Omega$ we find that the change
in $K_M$ induced by a change of morphology (dipolar vs multipolar cases) based on Figure~\ref{fig:aml} can be 
\begin{equation}
\frac{\frac{dJ}{dt}_{dip}}{\frac{dJ}{dt}_{mult}} =
\frac{K_{M_1}}{K_{M_0}} \sim 250,
\end{equation}
which is sufficient for the requirements of the MDM \citep{Brown14}.  The case $K_{M_0}$, referring to the regime in which the star is effectively decoupled from 
its wind, then corresponds to our multipolar topology, while $K_{M_1}$, 
referring to strong coupling, corresponds to development of a strong
dipolar component.  In Brown's approach the transition probability for this flip is mass
dependent and scales with the turnover time $\tau(M_{\star})$.  In our
framework, this is equivalent to a mass dependent timescale for the
evolution of the dynamo and magnetic morphology. 

The results presented here show that magnetic complexity can provide
almost an ``off-on" coupling switch between a star and its wind.  This
results from a combination of suppressed mass loss, a steeper
dependence of magnetic field with radial distance and more compact
Alfv\'en surface, and the shift of the dominant mass loss from
equatorial in the dipolar case to more spherically-symmetric as
complexity increases.  The MDM assumes an instantaneous irreversible
transition between dynamo modes.  In reality, there could be a period
of flip-flopping between different states of complexity,  perhaps
combined with a more gradual transition from higher order to lower
order states.  There is, however, currently no solid theoretical basis
for an underlying dynamo change at rotation periods of a few days that
might give rise to such changes.  The scenario could be tested
observationally through systematic examination of surface magnetic
field maps of stars on the $C$ and $I$ sequences, extending the
existing sample for stars on the $C$ sequence, and measurement of
their wind-driven mass loss rates.  While the apparently lower mass loss rates found for $\xi$~Boo~A and $\pi^1$~UMa by \cite{Linsky.Wood:14} are encouraging, with rotation periods of 6.3 and 5 days, respectively \citep{Donahue.etal:96,Maldonado.etal:10}
these stars have probably transitioned recently to the $I$ sequence.


\section{CONCLUSIONS}
\label{sec:Conclusions}

Using detailed 3D MHD models of a solar-like wind, we find
that mass loss and spin down rates get rapidly suppressed with 
increasing complexity of the stellar magnetic field. 
Higher order magnetic moments generate a magnetosphere with a
larger number of closed magnetic field lines that suppress mass loss.
For higher order magnetic moments, the spin-down rate is no longer a
simple function of the mass loss rate.  
Both closed and open field line regions become more homogeneously distributed over latitude for increasing
magnetic complexity.  
As a consequence, mass loss is no longer equatorially dominated as for
the dipolar case, and angular momentum loss is less efficient.
Furthermore, the steeper radial decrease of the magnetic field
magnitude for higher order multipoles results in a smaller Alfv\'en
surface and a shorter lever arm for magnetic braking. 
    
Echoing the MDM of \citet{Brown14}, we propose a new interpretation for the bimodal rotation distribution observed in young clusters:  Young solar-like stars have complex magnetic morphologies and lose angular
momentum in an inefficient way.   During this slow spin-down, magnetic complexity is eroded, precipitating a rapid transition from weak to strong wind coupling.  The very
rapid and drastic transition to the efficient spin down 
regime is caused by the very steep dependence of angular momentum loss on magnetic complexity found in this study.


\acknowledgments

CG and OC were supported by SI Grand Challenges grant ``Lessons from Mars: Are Habitable Atmospheres on Planets around M Dwarfs Viable?''.   OC was also supported by SI CGPS grant ``Can Exoplanets Around Red Dwarfs Maintain Habitable Atmospheres?''.  JJD was supported by NASA contract NAS8-03060 to the {\it Chandra X-ray Center}.
Numerical simulations were performed on the NASA HEC Pleiades system under award SMD-13-4526.



\end{document}